\documentclass[12pt]{article}
\usepackage{setspace}
\usepackage{amsmath,amssymb}
\usepackage{color}
\usepackage{hyperref}
\usepackage{array}
\usepackage{booktabs}
\numberwithin{equation}{section}
\usepackage{graphicx}
\usepackage{braket}
\usepackage[normalem]{ulem}

\usepackage{soul}

\newcommand{\beq}{\begin{equation}}
\newcommand{\eeq}[1]{\label{#1}\end{equation}}
\newcommand{\bea}{\begin{eqnarray}}
\newcommand{\eea}[1]{\label{#1}\end{eqnarray}}
\renewcommand{\Im}{{\rm Im}\,}
\renewcommand{\Re}{{\rm Re}\,}

\def\mS{\mathcal{S}}

\def\mM{\mathcal{M}}

\def\mM{\mathcal{M}}
\def\mS{\mathcal{S}}

\def\mA{\mathcal{A}}

\def\mI{\mathcal{I}}

\def\zb{\bar{z}}
\def\wb{\bar{w}}

\def\hb{\bar{h}}

\def\pa{\partial}

\def\g5{\gamma_5}

\def\b[#1]{\bold{#1}}
\def\bb[#1]{\overline{\bold{#1}}}
\def\bs[#1,#2]{\bold{#1}_{#2}}
\def\bbs[#1,#2]{\overline{\bold{#1}}_{#2}}

\def\s2{\sigma_2}
\def\ep{\epsilon}

\def\gammaflat{ \gamma_{z\zb}}
\def\gammaflatt{ \gamma^{z\zb}}

\def\Tsoft{T_{\text{soft}}}

\def\omegak{\omega_k}

\def\paz{\pa_z}
\def\pazU{\pa^z}
\def\pazb{\pa_{\zb}}
\def\pazbU{\pa^{\zb}}

\def\ketd[#1]{\ket{#1}_{\text{dressed}}}
\def\brad[#1]{\bra{#1}_{\text{dressed}}}
\def\ketas[#1]{\ket{#1}_{\text{Asymptotic}}}
\def\braas[#1]{\bra{#1}_{\text{Asymptotic}}}

\def\gammaflat{ \gamma_{z\zb}}

\def\Tsoft{T_{\text{soft}}}

\def\omegak{\omega_k}

\def\paz{\pa_z}
\def\pazU{\pa^z}
\def\pazb{\pa_{\zb}}
\def\pazbU{\pa^{\zb}}

\newcommand{\V}{\mathbf}

\def\bA{\bold A}
\def\bB{\bold B}

\def\bD{\boldsymbol{\nabla} }
\def\br{\bold r}
\def\bE{\bold E}
\def\bA{\bold A}
\def\bv{\bold v}
\def\bx{\bold x}
\def\bp{\bold p}

\def \ImCzz{\Im D_z ^2 C^{zz} }

\begin{document}
	\setlength{\topmargin}{-1cm} 
	\setlength{\oddsidemargin}{-0.25cm}
	\setlength{\evensidemargin}{0cm}

\begin{titlepage}
\hfill 
\begin{center}

\vskip 2 cm

{\LARGE \bf  Gravitational Wu-Yang Monopoles}

\vskip 2 cm

{Uri Kol  and Massimo Porrati}

\vskip .75 cm

{\em Center for Cosmology and Particle Physics\\
	Department of Physics, New York University \\
	726 Broadway, New York, NY 10003, USA}
	
	\vspace{12pt}

\end{center}

\vskip 1.25 cm

\begin{abstract}
\noindent

We show that in an asymptotically flat space where an S-Matrix can be defined, dual supertranslations 
leave all its matrix elements invariant and the Hilbert space of asymptotic states factorizes into distinct super-selection sectors, labeled by their dual supertranslation charges. These results suggest that dual supertranslation may be interpreted as a redundant gauge symmetry of asymptotically flat spacetimes.
This would allow to recast general relativity as a theory of diffeomorphisms possessing an additional asymptotic gauge symmetry.
We then use the conjectured dual supertranslation gauge symmetry to construct a gravitational equivalent of the Wu-Yang monopole solution. The metric describing the solution is defined using two overlapping patches on the celestial sphere. The solution is regular on each one of the patches separately and differentiable in the overlap region, where the two descriptions are identical by virtue of a dual supertranslation gauge transformation. Our construction provides an alternative to Misner's interpretation of the Taub-NUT metric. In particular, we find that using our approach the Taub-NUT metric can be made regular everywhere on the celestial sphere and at the same time it is devoid of closed timelike curves, provided that the  bound $\frac{m}{\ell} \leq  \sqrt{\frac{5}{27}}$ on the ratio of mass to NUT 
charge is obeyed.

\end{abstract}
\end{titlepage}
\newpage

\section{Introduction}

In 1976 Wu and Yang showed that the gauge potential of a  magnetic monopole is regular if it is interpreted as 
the connection of a nontrivial bundle. In their construction~\cite{Wu:1976ge} there are no Dirac strings, but the gauge potential is defined separately on two different patches of spacetime
\begin{equation}\label{WuYangPotentials}
\begin{cases}
\bA^N =+ g \frac{1-\cos \theta}{r \sin \theta} \hat{\varphi}, \qquad 0 \leq \theta  \leq \frac{\pi}{2} + \epsilon,
\\
\bA^S =- g \frac{1+\cos \theta}{r \sin \theta} \hat{\varphi}, \qquad  \frac{\pi}{2} - \ep  \leq \theta \leq \pi .
\end{cases}
\end{equation}
Both potentials result in the same magnetic field $\bB = \bD \times \bA = g \frac{\br}{r^3}$. In the overlap region the two potentials are equivalent up to a gauge transformation 
\begin{equation}
\bA ^S \rightarrow \bA ^S + \bD \lambda (\br)  =\bA ^N, \qquad  \lambda(\br) = 2 g\varphi.
\end{equation}
The potential $\bA^N$ would have contained a semi-infinite string singularity along the $\theta=\pi$ axis, but the singularity is outside the potential's domain of validity.
In the same way, the potential $\bA^S$ would have contained a semi-infinite string singularity along the $\theta=0$ axis, but here as well the singularity is outside the domain on which the potential is defined.
The solution is therefore devoid of any string singularities and it is described by a potential, which is differentiable in the overlap region where the two descriptions are equivalent up to a gauge transformation. An extensive review on magnetic monopoles can be found in \cite{Shnir}.

In this paper we construct a gravitational-magnetic monopole solution analogous to the Wu-Yang monopole in QED
by promoting the dual supertranslation symmetry of asymptotically flat spacetimes, discussed 
in~\cite{Kol:2019nkc,Godazgar:2018qpq}, to a gauge symmetry. Dual supertranslation is a symmetry that acts directly on the metric while leaving spacetime intact, whose action on the phase space cannot be reproduced by 
diffeomorphisms~\cite{Kol:2019nkc}. Moreover, the Bondi news, which is the gravitational analogue of the field strength in gauge theory and which is used as a measure of gravitational radiation, is inert under the action of dual supertranslations. These properties hint at the possibility that dual supertranslation may be a gauge symmetry of the metric.
We further argue in favor of dual supertranslation as a gauge symmetry of locally asymptotically flat spacetimes by showing that in theories where an S-Matrix can be defined, all S-Matrix elements are invariant under its action. Two metrics differing by the action of dual supertranslation can therefore be considered physically equivalent.

We provide additional evidence towards this picture by studying the magnetic zero mode of the metric. We show that the magnetic zero mode cannot be excited using a local stress-energy tensor and that the freedom in its definition corresponds to the dual supertranslation ambiguity.
These results suggest that dual supertranslation can be a redundant symmetry, in the sense that its conjugate excitations cannot be observed using local measurements.
Furthermore, we show that dual supertranslations factorize the Hilbert space of asymptotic states into distinct superselection sectors. Asymptotic states are classified according to their dual supertranslation charges and states from different superselection sectors cannot transform into each other.
These properties distinguish dual supertranslation from the standard physical supertranslation symmetry, and provide further evidence that it parametrizes a redundancy that can be interpreted as a gauge symmetry.

As a final piece of evidence, we study the geodesic equation of a massive particle. We show that in the leading post-Minkowskian approximation, the geodesic equation takes a form similar to the Lorentz force in electrodynamics. In particular, there is a magnetic force term $\bv \times \bB$, where the magnetic field is the curl of a vector potential that is given in terms of some metric components. We show that under dual supertranslations, the vector potential transforms as a gauge field and therefore the magnetic force, as well as the entire geodesic equation, is invariant under dual supertranslations.

We then proceed to construct the asymptotic form of a gravitational-magnetic monopole solution along the line of Wu-Yang by defining the asymptotic metric on two different patches separately.
The two metrics are perfectly regular in their own domains. 
We show that in the overlap region the two metrics are equivalent up to a dual supertranslation gauge transformation and therefore the solution is differentiable in this region.

Finally, we revisit the Taub-NUT spacetime as a non-linear gravitational-magnetic monopole solution.
Following the Wu-Yang mechanism we show that the Taub-NUT solution can be made regular, to leading order in the asymptotic expansion, by defining the metric using two different patches on the celestial sphere.
The metrics on the two patches are then shown to be equivalent in the overlap region up to a dual supertranslation transformation, so that the solution is differentiable in this region, when dual supertranslation is promoted
to a gauge symmetry.
Our approach provides an alternative to Misner's interpretation \cite{Misner:1963fr} who identified the two metrics in the overlap region by virtue of diffeomorphisms, which in turn implies that the time coordinate must be identified on the circle.
In Misner's construction both the metric \emph{and} the spacetime manifold are defined on two different patches and the identification results in the appearance of \emph{Closed Timelike Curves} (CTC).
Contrary to Misner's construction our spacetime manifold is globally defined and only the metric is written using two patches, which are identified in the overlap region by virtue of dual supertranslations. Therefore, the time coordinate remains non-compact in our construction. The azimuthal angle can still become timelike in certain regions of the solution, in which case CTC will appear since the angle is periodic. We remove this possibility by showing that the angle never becomes timelike provided that the following bound on the mass to NUT charge ratio is obeyed
 \begin{equation}\label{bound}
 \frac{m}{\ell}  \leq  \sqrt{\frac{5}{27}}.
 \end{equation}
 If the procedure outlined here could be extended to the entire spacetime, our solution would provide an explicit realization of a gravitational-magnetic monopole with neither string singularities nor CTC.

The paper is organized as follows. In section 2 we argue that dual supertranslation may be promoted to a gauge symmetry by 
showing that it acts trivially on all S-Matrix elements. In section 3 we study the magnetic zero mode and the superselection rule. In section 4 we study the geodesic equation of a massive particle and show that it is invariant under dual 
 supertranslations to leading order in the post-Minkowskian approximation.
In section 5 we construct the asymptotic form of a gravitational-magnetic monopole solution. In section 6 we revisit the 
Taub-NUT metric as a non-linear solution that describes a gravitational-magnetic monopole. Using dual supertranslations 
we show that the Taub-NUT solution can be made regular everywhere on the celestial sphere and that it is free of CTC,
 provided that the bound \eqref{bound} is satisfied. We end in section 7 with a discussion together with an alternative, more
conservative modification of the Taub-NUT solution, which contains a cosmic string but exhibits neither branch 
cuts nor closed timelike curves.

 \section{Dual Supertranslation as a Gauge Symmetry}

Asymptotically flat spacetimes can be expanded around future null infinity $\mI^+$ as follows
\begin{equation}\label{bondiMetric}
ds^2 = ds^2_{\text{Minkowski}} + \frac{2m_B}{r}du^2+r C_{zz} dz^2 + r C_{\zb \zb} d \zb^2 -2 U_z du dz -2 U_{\zb} du d\zb + \dots,
\end{equation}
where the radiative data $C_{zz}(u,z,\zb)$ is a function of all coordinates but $r$, $U_z = - \frac{1}{2} D^z C_{zz}$ and the dots indicate subleading terms in the expansion.
The flat Minkowski metric is given in terms of the round metric on the celestial sphere $\gammaflat$
\begin{equation}
ds^2_{\text{Minkowski}} = -du^2 -2dudr +2 r^2 \gammaflat dz d\zb , \qquad \gammaflat = \frac{2}{(1+z \zb)^2}.
\end{equation}
The function $m_B(u,z,\zb)$ denotes the Bondi mass aspect, while the Bondi news
\begin{equation}
N_{zz} (u,z,\zb) \equiv \pa_u C_{zz} ,
\end{equation}
is the analogue of the field strength in gauge theory and it is non-vanishing in the presence of gravitational radiation. At spatial infinity the following boundary condition is imposed
\beq
C_{zz} \Big|_{\mI^+_-} = D_z ^2 C(z,\zb),
\eeq{bg}
where $C(z,\zb)$ is a \emph{complex} function, commonly called the \emph{boundary graviton} \cite{Kol:2019nkc,Strominger:2013jfa,He:2014laa}. This boundary condition is necessary in order to get the correct action of supertranslations on the phase space.
The real and imaginary parts of the boundary graviton are the zero modes of the graviton field components $\Re D_z^2 C^{zz}$ and $\Im D_z^2 C^{zz}$, respectively. They were named ``electric" and ``magnetic" in \cite{Kol:2019nkc}, 
due to their spatial parity and in analogy to the corresponding field components in electrodynamics.

The metric \eqref{bondiMetric} is invariant under the BMS group, of which we will consider two elements; supertranslations, whose generator is given by
\begin{equation}\label{superT}
T(f) 
= \frac{1}{4\pi G}
\int_{\mI^+_-} d^2 z \gammaflat f(z,\zb) m_B \\
\end{equation}
and the recently discovered dual supertranslations, whose generator is
\begin{equation}\label{dualsuperT}
\mM(f)   
=  \frac{i}{16\pi G}
\int_{\mI^+_-}  d^2 z \,  \gammaflatt f(z,\zb) 
\Big(
D_{\zb}^2 C_{zz} -D_z^2 C_{\zb\zb}
\Big).
\end{equation}
Note that the supertranslation charge can be decomposed into soft and hard terms
\begin{equation}
\begin{aligned}
T(f) &= \Tsoft  (f) + T_{\text{hard}}(f),\\
\Tsoft  (f) &=- \frac{1}{16\pi G} \int_{\mI^+}  du \,  d^2z \, \gammaflatt  \,  \pa_u \Big(  D^2_{\zb} C_{zz}+ D^2_{z} C_{\zb \zb}    \Big)  f(z,\zb) , \\
T_{\text{hard}}(f) &= \frac{1}{4 \pi G} \int _{\mI^+} du \,  d^2z \, f(z,\zb) \,  \gammaflat  \, T_{uu},
\end{aligned}
\end{equation}
while the dual supertranslation charge receives contributions from soft gravitons only. 
The action of supertranslation and dual supertranslation on the phase space is given by
 \begin{equation}\label{ps2}
\begin{aligned}
\{T(f) , N_{zz }\} &= f \pa_u N_{zz},					   &  \qquad \{\mM(f)    , N_{zz }\} &=0,   \\
\{T(f) , C_{zz }\} &= f \pa_u C_{zz} -2D_z^2 f, 	&  \qquad \{\mM(f)    , C_{zz }\} &=-2i D_z^2 f,   \\
\{T(f) , \Re N\} &= 0, 													&  \qquad \{\mM(f)  ,  \Im N\} &= 0,   \\
\{T(f) , \Re C\} & = -2f, 											&  \qquad \{\mM(f)    , \Im C\} & = -2f, \\
\end{aligned}
\end{equation}
where $N(z,\zb)$ is a complex scalar field defined by $ \int_{-\infty}^{\infty} du N_{zz} (u,z,\zb)= D_z^2 N(z,\zb) $.
Eq.~\eqref{ps2} makes evident that the standard and dual supertranslation charges are conjugated to the real and imaginary parts of the boundary graviton, respectively.

Here we would like to emphasize two important properties of dual supertranslations:
\begin{itemize}
	\item The action of dual supertranslations on the phase space cannot be reproduced by using diffeomorphisms.
	\item The Bondi news, which is the gravitational analogue of the field strength in gauge theory, is invariant under the action of dual supertranslations.
\end{itemize}
These two properties suggest that dual supertranslation may describe a redundant gauge symmetry.
The last statement is supported by the fact that there is no Ward identity associated with dual supertranslations.
For comparison, standard supertranslations are associated with the following Ward identity
\begin{equation}\label{Ward}
 \bra{\text{out}} : T_{\text{soft}} (f) \mS: \ket{\text{in}}= \sum_{k} \left(E^{\text{in}}_k \, f(z^{\text{in}}_k) -E^{\text{out}}_k \, f(z^{\text{out}}_k) \right) 
  \bra{\text{out}}  \mS \ket{\text{in}},
\end{equation}
where the colons denote time ordering, $\mS$ is the S-Matrix, $z_k$ is the location of the $k$-th external particle on the celestial sphere and $E_k$ is its energy.
The Ward identity \eqref{Ward} was derived in \cite{Strominger:2013jfa,He:2014laa} and it follows from the conservation equation of BMS supertranslations and the decomposition of the charge into soft and hard parts. The left hand side of equation \eqref{Ward} describes the action of the soft part on S-Matrix elements while the right hand side describes the action of the hard part. This Ward identity relates S-Matrix elements with and without an additional insertion of a soft graviton and it implies that the action of soft supertranslation transformations on the vacuum is physical. On the other hand, the dual supertranslation charge receives contributions from soft gravitons only. As a consequence, the conservation equation in this case is
\begin{equation}
\bra{\text{out}} : \mM (f) \mS: \ket{\text{in}} =0 .
\end{equation}
It directly implies that the action of soft dual supertranslation transformations on the vacuum is trivial.
In the following, we will provide further evidence towards the statement that dual supertranslation can be promoted to a gauge symmetry by showing that all S-Matrix elements are invariant under its action.

\subsection*{Broken Symmetry versus Gauge Redundancy}

Both standard and dual supertranslations give rise, formally, to a freedom in the definition of the vacuum. In the following we study the effect of this ambiguity on S-Matrix elements. We will denote these S-Matrix elements generally by
\begin{equation}\label{amplitude}
\mA= \bra{0} \hat{\Psi}^\text{out} \mS  \hat{\Psi}^\text{in}  \ket{0},
\end{equation}
where $\hat{\Psi}^\text{in/out}$ denote matter operators acting on the incoming and outgoing vacuum states. Here we define the vacuum as the state that satisfies
\begin{equation}\label{vacuum}
a(\omega \hat{\V{x}}) \ket{0} = 0
\ ,
\end{equation}
where $a(\omega \hat{\V{x}}) $ is the graviton's annihilation operator.

When a symmetry is broken its charge does not exist; yet it is useful heuristically to imagine that an 
operator $Q(f)$, which commutes with the Hamiltonian, gives rise to a freedom in the definition of the vacuum
\beq
\ket{0} \rightarrow \ket{0} _{f} = \ket{0} +Q(f) \ket{0}
\eeq{broken}
and defines a family of zero energy states. Each one of these states $\ket{0}  _{f}$, parametrized by the continuous 
function $f$, could equally serve as the vacuum\footnote{This is why the charge in reality does not exist, as the
 vacua would be an uncountable family of normalizable orthogonal states, thereby 
 making the Hilbert space non-separable.}.
This ambiguity could be an observable one that gives rise to Goldstone bosons, or an unobservable gauge redundancy.
In the first case different states in the family of vacua are physically distinguishable while in the second case they are indistinguishable.
To determine whether the ambiguity is physical or not we will work with operators defined schematically as
$\int_{I^+} J(u,z,\bar{z})\exp(i\omega u)$ and take the limit $\omega\rightarrow 0$. The operators and 
amplitudes containing them are well defined in the limit $\omega \rightarrow 0$; this is what justifies formal
manipulations done at $\omega=0$.  We will compute the S-Matrix element between states in the family
\begin{equation}
\mA _f = \bra{0}_f \hat{\Psi}^\text{out} \mS  \hat{\Psi}^\text{in}  \ket{0},
\end{equation}
and check if they depend on the transformation parameter $f$ that characterizes the freedom in the definition of the vacuum.
A related discussion was recently made in \cite{Choi:2019sjs}, where the authors have studied properties of dressing operators.

\subsection*{Standard Supertranslations}

We start by studying standard supertranslations, whose generator $T(f)$ can be used to define a family of zero energy states for any function $f(z,\zb)$
\begin{equation}
T(f) \ket{0} = \Tsoft (f) \ket{0} .
\end{equation}
Note that the hard part inside $T(f)$ annihilates the vacuum. To compute $\mA_f$ in terms of $\mA$ we will use Weinberg's soft theorem, which relates the amplitude \eqref{amplitude} with an additional soft graviton insertion to the original amplitude
\begin{equation}\label{softTheorem}
\begin{aligned}
\lim_{\omegak \rightarrow 0} \omegak \mA_{k,\text{soft}}&=
\lim_{\omegak \rightarrow 0} \omegak \bra{0} a^r(k)\hat{\Psi}^\text{out} \mS \hat{\Psi}^\text{in}  \ket{0}\\
&=
\frac{\kappa}{2} \lim_{\omega_k \rightarrow 0}  \omegak
\left(   \sum_i \eta_i \frac{p_i^{\mu}p_i^{\nu}}{p_i \cdot k} \right)\ep^r_{\mu\nu}(k) \mA.
\end{aligned}
\end{equation}
$\omega_k,k$ and $r$ are the energy, four-momentum and polarization of the soft graviton, respectively.
The sum is over all external particles whose momenta are given by $p^{\mu}_i$.
$\eta_i=+1$ for an outgoing particle and $\eta_i=-1$ for an incoming particle.
$\kappa$ is related to Newton's constant by $\kappa^2 = 32 \pi G$.
The graviton's annihilation operator can be decomposed in terms of polarization modes
\begin{equation}
a_{\mu\nu} (k) =\sum _{r=\pm} \ep^{r*}_{\mu\nu} (k) a_r(k).
\end{equation}
The transverse-traceless components of the polarization tensor can be decomposed as follows
\begin{eqnarray}
\ep^{\pm}_{\mu\nu} (k) = \ep^{\pm}_{\mu}(k)\ep^{\pm}_{\nu}(k).
\end{eqnarray}

Applying the supertranslation generator to the vacum
\begin{equation}\label{superSoft}
\bra{0} T(f) =
\lim_{\omega_k \rightarrow 0}\frac{ \omega_k }{4 \pi  \kappa}
\int  d^2z \bra{0} \left[
a_-(\omega_k \hat{x}) D_z^2 f
+a_+ (\omega_k \hat{x}) D_{\zb}^2 f
\right] 
\end{equation}
we can now compute
\begin{equation}\label{derSuper}
\begin{aligned}
\bra{0} T(f) \hat{\Psi}^\text{out} \mS  \hat{\Psi}^\text{in}  \ket{0} &=
\lim_{\omega_k \rightarrow 0}\frac{ \omega_k }{4 \pi  \kappa}
\int  d^2z  \left[ 
\bra{0}a_-(\omega_k \hat{x})\hat{\Psi}^\text{out}  \mS  \hat{\Psi}^\text{in}  \ket{0}   D_z^2 f
\right. \\ &  \left. 
\qquad \qquad  \qquad \quad  \; \; \: +\bra{0}a_+ (\omega_k \hat{x}) \hat{\Psi}^\text{out} \mS \hat{\Psi}^\text{in}  \ket{0}  D_{\zb}^2 f
\right] .
\end{aligned}
\end{equation}
Using the soft theorem \eqref{softTheorem} the last expression takes the form
\begin{equation}
\begin{aligned}
&\bra{0} T(f) \hat{\Psi}^\text{out} \mS  \hat{\Psi}^\text{in}  \ket{0} 
=
\lim_{\omega_k \rightarrow 0} \mA \times \frac{ \omega_k }{8 \pi  }
\int  d^2z  
\Big( \ep^-_{\mu\nu}(k)  D_z^2 f+\ep^+_{\mu\nu}(k)  D_{\zb}^2 f   \Big)
\sum_i  \left( 
\eta_i \frac{p_i^{\mu}p_i^{\nu}}{p_i \cdot k}
\right)
\\
&
\qquad \qquad \quad
=
\lim_{\omega_k \rightarrow 0}  \mA \times  \frac{ \omega_k }{8 \pi  }
\int  d^2z  \left[ 
D_z^2 f \sum_i  \left(\eta_i \frac{ (p_i\cdot \ep^-)^2 }{p_i \cdot k} \right)
+
D_{\zb}^2 f \sum_i  \left(\eta_i \frac{(p_i\cdot \ep^+)^2 }{p_i \cdot k}\right)
\right] .
\end{aligned}
\end{equation}
Integrating by parts gives\footnote{The action of the operator $D_z^2$ on a scalar function is given by $D_z^2=\gammaflat \paz \gamma^{z\zb} \paz=\gammaflat \paz \pa^{\zb}$.}
\begin{equation}\label{derivation}
\begin{aligned}
\bra{0} T(f) \hat{\Psi}^\text{out} \mS  \hat{\Psi}^\text{in}  \ket{0} 
&=
\lim_{\omega_k \rightarrow 0}  \mA \times \frac{ \omega_k }{8 \pi  }
\int  d^2z f (z,\zb)  \left[ 
\paz \pa^{\zb} \left(   \gammaflat \sum_i \eta_i  \frac{ (p_i\cdot \ep^-)^2 }{p_i \cdot k} \right)
\right. \\ &  \left. 
\qquad \qquad  \qquad \qquad \qquad \qquad \;  
+
\pazb \pazU \left( \gammaflat \sum_i \eta_i  \frac{ (p_i\cdot \ep^+)^2 }{p_i \cdot k} \right)
\right] .
\end{aligned}
\end{equation}
We now use the following identity \cite{Choi:2017bna}
\begin{equation}\label{identity}
\paz \pazbU    \left(    \gammaflat  \frac{ (p \cdot \ep^-)^2 }{\hat{k}_{z,\zb}\cdot p}   \right)   
=\pazb \pazU    \left(    \gammaflat  \frac{(p \cdot \ep^+)^2}{\hat{k}_{z,\zb}\cdot p}   \right)  
=
\frac{1}{2} \gammaflat \frac{p^4}{(p\cdot \hat{k}_{z,\zb} )^3} 
=
\frac{1}{2} \gammaflat \frac{m^4}{(p\cdot \hat{k}_{z,\zb} )^3} 
\end{equation}
to show that the two terms in \eqref{derivation} are equal and add up to give
\begin{equation}
\begin{aligned}
\bra{0} T(f) \hat{\Psi}^\text{out} \mS  \hat{\Psi}^\text{in}  \ket{0} 
&=
\mA \times \frac{ 1 }{8 \pi  }
\int  d^2z  \gammaflat  \, f(z,\zb)
\sum_i  \left(
\eta_i   \frac{m_i^4}{(p_i\cdot \hat{k}_{z,\zb} )^3} 
\right),
\end{aligned}
\end{equation}
where $m_i$ is the mass of the $i$th external particle.
As a consequence, the result for the supertranslated amplitude $\mA_f$ is given by
\begin{equation}
\mA_f =\mA \times  \left(1+ 
\frac{ 1 }{8 \pi  }
\int  d^2z  \gammaflat  \, f(z,\zb)
\sum_i  
\eta_i   \frac{m_i^4}{(p_i\cdot \hat{k}_{z,\zb} )^3} 
\right).
\end{equation}
In particular we notice that $\mA_f \neq \mA$ and so the action of supertranslations on the vacuum can be physically observable. Scattering amplitudes between different vacuum states in the family $\ket{0}_f$ are different because
they contain a  different number of soft gravitons.

\subsection*{Dual Supertranslations}

We now repeat the same derivation for dual supertranslations, whose action on the vacuum is given by
\begin{equation}\label{dualSuperSoft}
\bra{0} \mM(f) =
\lim_{ \omega_k \rightarrow 0}\frac{ i \omega_k }{4 \pi  \kappa}
\int  d^2z \bra{0} \left[
a_-(\omega_k \hat{x}) D_z^2 f
-a_+ (\omega_k \hat{x}) D_{\zb}^2 f
\right] 
\end{equation}
Besides an overall factor of $i$, this formula differs from the corresponding expression for standard supertranslations 
because of a crucial relative minus sign between the two terms within brackets.
We can now compute the transition amplitude 
\begin{equation}
\begin{aligned}
\bra{0} \mM(f) \hat{\Psi}^\text{out} \mS  \hat{\Psi}^\text{in}  \ket{0} &=
\lim_{\omega_k \rightarrow 0}\frac{ i  \omega_k }{4 \pi  \kappa}
\int  d^2z  \left[ 
\bra{0}a_-(\omega_k \hat{x})\hat{\Psi}^\text{out}  \mS  \hat{\Psi}^\text{in}  \ket{0}   D_z^2 f
\right. \\ &  \left. 
\qquad \qquad  \qquad \quad  \; \; \: -\bra{0}a_+ (\omega_k \hat{x}) \hat{\Psi}^\text{out} \mS \hat{\Psi}^\text{in}  \ket{0}  D_{\zb}^2 f
\right] ,
\end{aligned}
\end{equation}
which is very similar to the result for supertranslations \eqref{derSuper} except for the relative minus sign between the two terms.
We have seen that these two terms are equal by virtue of the identity \eqref{identity}. In the case of standard supertranslations, they added up, but in the present case they will precisely cancel each other to give
\begin{equation}\label{SuperSelectionRule}
\bra{0} \mM(f) \hat{\Psi}^\text{out} \mS  \hat{\Psi}^\text{in}  \ket{0} =0.
\end{equation}
This is one of our main results in this paper as it implies that S-Matrix elements are independent of dual supertranslation transformations of the vacuum
\begin{equation}
\mA _f = \mA \, .
\end{equation}
We conclude that dual supertranslation can be interpreted as an unobservable gauge symmetry.

\section{Superselection Rule}

The result \eqref{SuperSelectionRule} is a superselection rule. Asymptotic states can be classified into distinct superselection sectors according to their dual supertranslation charges and the superselection rule \eqref{SuperSelectionRule} implies that states from different sectors cannot transform into each other.

In this section we describe the dynamical origin of the superselection rule by studying the two zero modes of the asymptotic graviton $C_{zz}(u,z,\zb)$, which are the $u$-independent solutions. We start by reviewing the dynamics of the electric zero mode, which is constrained by the $uu$ component of the Einstein equations $G_{uu}=8\pi T_{uu}^M$
\begin{equation}\label{uuEquation}
	\begin{aligned}
		\pa_u m_B &= \frac{1}{4}\pa_u \left(D_z^2 C^{zz}+D_{\zb}^2 C^{\zb\zb}\right) - T_{uu} ,\\
		T_{uu} &=  4\pi G \lim_{r \rightarrow \infty } \left(r^2 T^M_{uu}\right) + \frac{1}{4} N_{zz}N^{zz}.
	\end{aligned}
\end{equation}
The solution for the electric component of the asymptotic metric is
\begin{equation}\label{electricMode}
	\Re \left[  D_z^2 C^{zz}(u,z,\zb) \right]=2 \Delta m_B(u,z,\zb)+ 2 \int_{-\infty}^u du' \, T_{uu}(u',z,\zb) ,
\end{equation}
where $ \Delta m_B(u,z,\zb) =  \Delta m_B(u,z,\zb)- \Delta m_B(-\infty,z,\zb)$.
In particular, the zero mode of $\Re \left(   D_z^2 C^{zz} \right)$ can be excited by a localized source, namely using a stress tensor that is non-zero only in a finite interval $u_i<u<u_f$ and which vanishes at spatial infinity.

The magnetic zero mode, on the other hand, is constrained by the $uz$ component of the Einstein equations $G_{uz}=8\pi T^M_{uz}$
\begin{equation}\label{uzEquation}
	\begin{aligned}
		\pa_u N_z &= \frac{1}{4}\paz  \left(D_z^2 C^{zz} - D_{\zb}^2 C^{\zb\zb}\right)  -u \pa_u \paz m_B  - T_{uz} ,\\
		T_{uz} &= 8 \pi G \lim_{r \rightarrow \infty } \left(r^2 T^M_{uz}\right) - \frac{1}{4} \paz \left( C_{zz}N^{zz}\right) -\frac{1}{2}C_{zz}D_z N^{zz}.
	\end{aligned}
\end{equation}
The solution for the magnetic component of the asymptotic metric is then given by
\begin{equation}\label{magEq}
	\ImCzz 
	=
	-\frac{1}{2} \ell +
	 i  \int d^2 w  \, 
	\Big(
	\pa_{\wb}G(z;w) \left( \pa_u N_{w} +T_{u w}\right)
	-
	\pa_{w}G(z;w) \left(  \pa_u N_{\wb} +T_{u \wb}\right)
	\Big),
\end{equation}
where $G(z;w)$ is the Green's function (see appendix \ref{MagAppendix} for more details).
The first term in the equation above is the magnetic zero mode, which is constant on the sphere and independent of $u$ (the factor of $-1/2$ is for notation only). The response of the magnetic mode to the stress tensor is instantaneous and does not involve integration over $u$ as in the solution for the electric mode \eqref{electricMode}. We therefore conclude that the magnetic zero mode cannot be excited by a local stress-energy tensor.

To further examine the magnetic zero mode we now explore the solutions of the following equation
\begin{equation}\label{spaceSol}
	\ImCzz \equiv \frac{D_z^2 C^{zz} - D_{\zb}^2 C^{\zb\zb}}{2i} = -\frac{1}{2}  \ell .
\end{equation}
This equation can be solved using the boundary graviton $C_{zz}=D_z^2 C(z,\zb)$. It follows that the imaginary part of the boundary graviton obeys
\begin{equation}
D_z^2 D_{\zb}^2 \,  \Im C(z,\zb) = \frac{1}{2}  \ell \gammaflat ^2,
\end{equation}
while its real part is left unconstrained by \eqref{spaceSol}.
The solution for the imaginary part of $C$ is given by
\begin{equation}\label{solTN}
\Im C (z,\zb) =   \ell \log \frac{1+|z|^2}{|z|}.
\end{equation}
The solution \eqref{solTN} describes the Taub-NUT metric with an infinite string singularity at $z=0,\infty$ (corresponding to $\theta=0,\pi$), see \cite{Kol:2019nkc} for more details. However, this solution is not unique since it can be shifted by a solution of the homogeneous equation
\begin{equation}\label{InvCond}
D_z^2 D_{\zb}^2 \,  f(z,\zb) = 0.
\end{equation} 
The solution to \eqref{InvCond} is given by a combination of holomorphic and anti-holomorphic functions on the sphere
\begin{equation}\label{ambiguity}
f (z,\zb) =  h(z) + \hb (\zb)  .
\end{equation}
For example, one can use combinations of the following functions
\begin{equation}
h(z) =  \log z , \qquad \hb(\zb)  =  \log \zb ,
\end{equation}
to change the solution \eqref{solTN} into
\begin{equation}
\Im C (z,\zb) =  \ell \log (1+|z|^2) \qquad  \text{or} \qquad    \ell \log (1+|z|^{-2}),
\end{equation}
which describe semi-infinite strings pointing in opposite directions \cite{Kol:2019nkc}.
The freedom described by the function $f(z,\zb)$ corresponds to a dual supertranslation transformation.
The condition \eqref{InvCond} imposes that the dual supertranslation charge stays invariant under a dual supertranslation transformation
\begin{equation}
\mM(f_2): \, \mM(f_1) \rightarrow 
\mM(f_1)
-\frac{1}{8 \pi G} \int_{\mI^+_-} d^2 z \, \gamma^{z\zb} \, f_1(z,\zb)  \, D_z^2 D_{\zb}^2 f_2(z,\zb)=\mM(f_1).
\end{equation}

We conclude this section by summarizing the two main results that we derived here:
\begin{itemize}
	\item The magnetic zero mode cannot be excited by a local stress-energy tensor.
	\item The freedom in the definition of the magnetic zero mode corresponds to the dual supertranslation ambiguity.
\end{itemize}
In particular, the dual supertranslation charge \eqref{dualsuperT} can never change through local stress tensor fluctuations since it is given by an integral over the magnetic zero mode. These results imply that dual supertranslations factorize the Hilbert space of asymptotic states into distinct superselection sectors.
Asymptotic states are therefore classified according to their dual supertranslation charge and states from different superselection sectors cannot transform into each other. These properties distinguish dual supertranslation from the standard physical supertranslation symmetry, and provide further evidence that it may be a redundant gauge symmetry.

\section{The Geodesic Equations}

In the following we wish to provide further evidence towards the picture advocated in the previous sections, according to 
which dual supertranslations describe an unphysical ambiguity. This evidence comes from studying the geodesic equations of a massive particle 
moving in the background \eqref{bondiMetric}. Since dual supertranslations do not act on the matter degrees of 
freedom, a necessary condition that they have to satisfy to be gauge symmetries is that the geodesic 
equations be not only covariant but also {\em invariant} under their action.

To discuss the geodesic equation we use a Cartesian system of coordinates $\left(t, \bx\right)$.
The four-momentum of a massive test particle
\begin{equation}
p^{\mu}= m _0 u^{\mu}
\end{equation}
is given in terms of its four-velocity, which takes the following form in Cartesian coordinates
\begin{equation}
u^{\mu}=\frac{dx^{\mu}}{d\tau} = \gamma \left(1,\bv\right) ,
\end{equation}
where $m_0$ is its mass.
The Lorentz factor is
\begin{equation}
\gamma = \frac{dt}{d \tau} = \left( 1-\bv\cdot \bv\right)^{-1/2}.
\end{equation}
The geodesic equation of the test particle is given by
\begin{equation}
\frac{d^2 x^{\mu}}{d\tau^2} = - \Gamma ^{\mu}_{\nu \rho} \frac{dx^{\nu}}{d\tau}\frac{dx^{\rho}}{d\tau}.
\end{equation}
The spatial components of this equation take the form
\begin{equation}
\frac{d (	\gamma v^{i}) }{d\tau} = - \gamma ^2 \left(  \Gamma ^{i}_{00} 
+2 \Gamma ^{i}_{0j} v^j
+ \Gamma ^{i}_{j k} v^j v^k \right)  ,
.
\end{equation}
where the indices $(i,j,k)$ run over the three Cartesian coordinates $\bx$ and we have used that $\frac{dx^0}{d\tau} = \gamma$. Namely,
\begin{equation}\label{geoEq0}
-\gamma^{-2} \frac{d  (	\gamma v^{i})  }{d\tau} =  \Gamma ^{i}_{00} 
+2 \Gamma ^{i}_{0j} v^j
+ \Gamma ^{i}_{j k} v^j v^k 
.
\end{equation}
Explicitly, the Christoffel symbols are given by
\begin{equation}
\begin{aligned}
\Gamma ^i_{00} &= \frac{1}{2} \eta^{ij} \left(2 \pa_{t} h_{0j} - \pa_j h_{00} \right) \\
\Gamma ^i_{0j} &= \frac{1}{2} \eta^{ik} \left(  \pa_t h_{jk} +\pa_j h_{0k} - \pa_k h_{0j} \right) \\
\Gamma ^i_{jk} &= \frac{1}{2} \eta^{in} \left(  \pa_j h_{kn}  + \pa_k h_{jn} - \pa_n h_{jk} \right) 
\end{aligned}
\end{equation}
where $g_{\mu\nu} = \eta_{\mu\nu} + h_{\mu\nu}$ and $\eta_{ij} $ is the flat spatial metric.

We now want to express the Christoffel symbols in the geodesic equation \eqref{geoEq0} in terms of the metric components of \eqref{bondiMetric}.
To simplify the analysis we perform the following change of coordinates
\begin{equation}\label{changeCoord}
\begin{aligned}
u & \rightarrow u+ F(r), \qquad \text{with}\qquad  F'(r)=- \frac{2m_B}{r} \left(1- \frac{2m_B}{r} \right)^{-1}  ,\\
z & \rightarrow z- \frac{1}{r} V^z ,\\
\zb & \rightarrow \zb- \frac{1}{r} V^{\zb}  .
\end{aligned}
\end{equation}
with functions $V^z,V^{\zb}$ to be determined soon.
It will be easiest to work with the system of coordinates $\left(t,r,z,\zb\right)$.
The geodesic equation \eqref{geoEq0} takes the same form in this system of coordinates except that the spatial indices now run over $(i,j,k)\sim (r,z,\zb)$ and the metric is expanded around the flat metric
\begin{equation}
g_{\mu\nu} = \eta_{\mu\nu} + h_{\mu\nu},
\end{equation}
where
\begin{equation}
\eta_{\mu\nu} =
\left(
\begin{tabular}{cccc}
$-1$ & $0$ & $0$ & $0$  \\
$0$ & $+1$ & $0$ & $0$  \\
$0$ & $0$ & $0$ & $r^2 \gammaflat $  \\
$0$ & $0$ & $r^2 \gammaflat $ & $0$  
\end{tabular}
\right).
\end{equation}
After the change of coordinates \eqref{changeCoord} and switching to the coordinates $\left(t,r,z,\zb\right)$, the metric \eqref{bondiMetric} takes the form
\begin{equation}\label{NPMmetric}
\begin{aligned}
h_{00}&= \frac{2m_B}{r}, \\
h_{0i} &= \left(
0 , -U_z , - U_{\zb}
\right),\\
h_{ij} &=   \left(
\begin{tabular}{ccc}
$\frac{2m_B}{r}$ & $U_z+V_z$ & $U_{\zb}+V_{\zb}$   \\
$U_z+V_z$ & $r \left( C_{zz} -2D_z V_z\right)$ & $-r \, \pa_{(z}V_{\zb)}$   \\
$U_{\zb}+V_{\zb}$ & $-r \, \pa_{(z}V_{\zb)}$ & $r \left( C_{\zb\zb} -2D_{\zb} V_{\zb}\right)$  
\end{tabular}
\right),
\end{aligned}
\end{equation}
where $\pa_{(z}V_{\zb)} \equiv \paz V_{\zb}+\pazb V_z$.
To simplify  $h_{ij}$ further we now choose the functions $V^z,V^{\zb}$	such that\footnote{Another simplifying choice could be $V_z = - U_z$.}
\begin{equation}
\begin{aligned}
C_{zz} &= 2D_z V_z, \\
C_{\zb\zb} &= 2D_{\zb} V_{\zb} .
\end{aligned}
\end{equation}
We can invert these relations using the Green's function
\begin{equation}
G(z;w) = \frac{1}{2\pi} \log |z-w|^2 ,
\end{equation}
which satisfies
\begin{equation}
\paz \pazb G(z;w) = \delta^2 (z-w),
\end{equation}
such that
\begin{equation}
\begin{aligned}
V^z &= \frac{1}{2} \int d^2w \, G(z;w) \, \pa_w \left(\gamma^{w \wb} C_{\wb\wb}\right), \\
V^{\zb} &= \frac{1}{2} \int d^2w \, G(z;w) \, \pa_{\wb} \left(\gamma^{w \wb} C_{ww}\right).
\end{aligned}
\end{equation}
Now under dual supertranslations
\begin{equation}
\begin{aligned}
C_{zz} & \rightarrow C_{zz} + i D_z^2 f, \\
C_{\zb\zb} & \rightarrow C_{\zb\zb} - i D_{\zb}^2 f,\\
m_B & \rightarrow m_B,
\end{aligned}
\end{equation}
where $f(z,\zb)$ is a real function, the functions $V^z,V^{\zb}$ transform as
\begin{equation}
\begin{aligned}
V^z& \rightarrow V^z - \frac{i}{2} \pa^z f, \\
V^{\zb}& \rightarrow V^{\zb} + \frac{i}{2} \pa^{\zb} f.
\end{aligned}
\end{equation}
Therefore 
\begin{equation}
\pa_{(z}V_{\zb)}   \rightarrow \pa_{(z}V_{\zb)}
\end{equation}
is invariant under dual supertranslations, and 
\begin{equation}
U_z + V_z   \rightarrow U_z + V_z - \frac{i}{2} \pa_z D^z D_z f.
\end{equation}
On the space of solutions with a non-zero dual supertranslation charge given in eq.~\eqref{spaceSol}, the imaginary part of the boundary graviton obeys the following equation
\begin{equation}
D^z D_z \Im C  = -2 \ell
\end{equation}
where $\ell$ is  constant. Solutions to the equation above are defined up to gauge transformations that obey
\begin{equation}
D^z D_z f = 0.
\end{equation}
Therefore  the combinations $U_z + V_z$ and $U_{\zb} + V_{\zb}$ are in fact invariant under dual supertranslation. Since the Bondi mass $m_B$ is also invariant, we therefore see that $h_{ij}$ is completely invariant under dual supertranslations.  $h_{00}$ is also invariant. The only metric components that are not invariant under dual supertranslations are $h_{0i}$.

We can now write the geodesic equation in compact form.
Recall that $m_B,U_{z}$ and $U_{\zb}$ are all functions of $(t,z,\zb)$.
We define
\begin{equation}\label{potentials}
\begin{aligned}
\phi & \equiv - \frac{1}{2}h_{00} , \\
A_i &\equiv +  h_{0i} ,
\end{aligned}
\end{equation}
in terms of which the geodesic equation takes the form
\begin{equation}\label{geoEq}
-\gamma^{-2} \frac{d  (	\gamma v^{i})   }{d\tau} =
\Big(
\bD \phi 
-\bv \times (\bD \times \bA ) 
+ \pa_t \bA
\Big)^i
+\eta ^{ik}v^j \pa_t h_{jk}
+\Gamma^i_{jk}v^jv^k
.
\end{equation}
The first term in the geodesic equation depends only on the Bondi mass; it is therefore invariant under dual supertranslations.
The last two terms are invariant as well, because they depend only on $h_{ij}$.
The term $\pa_t \bA$ is invariant because of the time derivative.
The only term that depends on $\Im C(z,\zb)$ is the magnetic force $\bv\times (\bD \times \bA)$, which we will now study to check whether it is invariant under dual supertranslations or not.

Let us write
\begin{equation}
C_{zz}(u,z,\zb) = C_{zz}'(u,z,\zb) +D_z^2 C(z,\zb),
\end{equation}
where $C(z,\zb)$ is a complex function. Under dual supertranslation $C_{zz}'(u,z,\zb)$ remains inert while $C(z,\zb)$ transforms as
\begin{equation}
C(z,\zb) \rightarrow C(z,\zb) + i f(z,\zb),
\end{equation}
where $f(z,\zb)$ is a real function. Under dual supertranslation the vector $U_z$ then transforms as
\begin{equation}
\begin{aligned}
U_z &\rightarrow U_z - i \frac{1}{2} D^z D^2_z f 
=
U_z +\pa_z \chi ,
\\
U_{\zb} &\rightarrow U_{\zb} + i \frac{1}{2} D^{\zb} D^2_{\zb} f 
=
U_{\zb} +\pa_{\zb} \chi ^*  ,
\end{aligned}
\end{equation}
where
\begin{equation}
\chi = -\frac{i}{2} \left(f + D^z D_z f \right)=-\frac{i}{2} f.
\end{equation}
In the last equality we have used that $D^z D_z f=0$ for dual supertranslation gauge transformations.
We can write the transformation law of $\bA$ under dual supertranslation as
\begin{equation}
\bA \rightarrow\bA +	 \bD \Lambda
\end{equation}
where
\begin{equation}
\Lambda = i ( h (z) - \hb (\zb) ).
\end{equation}
We see that $\bA$ transforms as an Abelian gauge field under dual supertranslations, therefore the magnetic force in the geodesic equation \eqref{geoEq} stays invariant. The conclusion is that the geodesic equation is invariant under dual supertranslations.

The geodesic equation takes a form similar to that of the relativistic electromagnetic Lorentz force
\begin{equation}
\frac{d \bp}{dt} = q \left( \bE + \bv \times \bB \right)
\end{equation}
acting on a particle of charge $q$, mass $m_0$ and momentum $\bp = \gamma m_0 \bv$, where
\begin{equation}
\begin{aligned}
\bE &= - \bD \phi - \pa_t \bA, \\
\bB &= \bD \times \bA.
\end{aligned}
\end{equation}
Using the definitions \eqref{potentials} the geodesic equation \eqref{geoEq} takes the form
\begin{equation}\label{geoEq2}
\frac{d \bp}{dt} = m_0 \gamma \left( \bE + \bv \times \bB - v_j \pa_t h^{ij} - \Gamma ^i _{jk} v^j v^k \right).
\end{equation}
We see that $q$ is replaced with $m_0$ and that there is an additional overall factor of $\gamma$ with respect to the electromagnetic force.
In the Taub-NUT background, the last two terms in \eqref{geoEq2} were found to give rise to transient forces \cite{Huang:2019cja}.
It would be interesting to generalize the results of \cite{Huang:2019cja} to the post-Minkowskian metric \eqref{bondiMetric} and provide a physical understanding for the last two terms in the general case.
Let us conclude by summarizing that in the geodesic equation both the electric and magnetic fields, as well as the two last terms, are invariant under dual supertranslation
\begin{equation}
\begin{aligned}
\bE & \rightarrow \bE, \\
\bB &\rightarrow \bB + \bD \times \bD \Lambda = \bB.
\end{aligned}
\end{equation}

Finally, we would like to add that the contribution of the vector field $\bA$ to the Ricci scalar, in the linearized approximation, is given by
\begin{equation}
R = -2 \pa_t \left(\bD \cdot \bA \right) + \dots ,
\end{equation}
where the dots represent terms that depend on $h_{00}$ or $h_{ij}$, but not on $h_{0i}=A_i$. The Ricci scalar is therefore invariant under dual supertranslations.

\section{Gravitational-Magnetic Monopole}

In the previous sections we provided evidence showing that dual supertranslation can be regarded as a redundant gauge symmetry. As a result, two metrics that differ by the action of dual supertranslation are physically indistinguishable. For example, suppose that we apply dual supertranslation to the Minkowski vacuum. The resulting metric is given by \eqref{bondiMetric} with
\begin{equation}\label{monopoleData}
\begin{aligned}
m_B &=0, \\
C_{zz} &= D_z^2 C(z,\zb),
\end{aligned}
\end{equation}
where $C(z,\zb)$ is an imaginary function which is globally defined on the sphere.
This metric is a pure gauge, namely it cannot be distinguished from the Minkowski vacuum.

At this point we could ask a rather obvious question: if dual supertranslation is a gauge symmetry, whose only 
effect is to transform the redundant degree of freedom 
$\Im  C(z,\zb)$ [see eq.~(\ref{bg})], why can't we set $\Im C(z,\zb)=0$ in the first place and keep only standard
supertranslations  as true asymptotic symmetries?  We could, of course, but by doing it 
we would be throwing away solutions where the
metric is a nontrivial section of a bundle on the celestial sphere, with transition functions defined by dual supertranslations. 
An example is given by a monopole solution, where, following the Wu-Yang procedure, we define the metric using \eqref{bondiMetric} and \eqref{monopoleData}, but with a boundary graviton which is defined on two overlapping patches separately
\begin{equation}\label{Cs}
\begin{cases}
C_N =-4 i \ell \log \left(1+|z|^{+2}\right) =  8 i \ell \log \cos \frac{\theta}{2},    \qquad 0 \leq \theta  \leq \frac{\pi}{2} + \epsilon,
\\
C_S =-4 i \ell \log \left(1+|z|^{-2}\right) =  8 i \ell \log \sin \frac{\theta}{2}, \qquad  \frac{\pi}{2} - \ep  \leq \theta \leq \pi .
\end{cases}
\end{equation}
This solution is distinguishable from the Minkowski vacuum because it cannot be obtained by acting on the later with a globally defined, regular, dual supertranslation transformation.
The expression we chose for $C_N$ describes a semi-infinite gravito-magnetic string located along the $\theta=\pi$ axis \cite{Kol:2019nkc}.
Similarly, $C_S$ describes a semi-infinite gravito-magnetic string located along the $\theta=0$ axis.
In both cases the singularity occurs outside the domain on which the patch is defined and therefore the metric is perfectly regular.
However, we still need to make sure that the solution is also differentiable in the overlap region $\frac{\pi}{2} - \ep  \leq \theta \leq \frac{\pi}{2} + \ep$. We do so by showing that in the overlap region the two metrics are identical up to a dual supertranslation transformation $\mM(f)$, with a parameter
\begin{equation}\label{trans}
f(z,\zb) = C_N (z,\zb)  - C_S (z,\zb)  = -8 i \ell \log  |z|= -8 i \ell \log  \tan \frac{\theta}{2} .
\end{equation}
This gauge transformation diverges at the two poles but is perfectly regular in the overlap region.
Dual supertranslations do not act on the spacetime coordinates and therefore the two metrics are identical in the overlap region up to a gauge transformation that acts directly on the metric. The global dual supertranslation charge of the proposed solution receives two equal contributions, one from each patch
\begin{equation}
\mM_{\text{global}} = \mM_{\text{global}}^{(N)}+ \mM_{\text{global}}^{(S)} = \frac{\ell}{2G}+\frac{\ell}{2G} = \frac{\ell}{G} ,
\end{equation}
so it is precisely equal to the dual supertranslation charge of the Taub-NUT metric.

The construction above describes the asymptotic form of a gravito-magnetic monopole and can be extended into the bulk by solving the Einstein equations order by by order. If no further asymptotic data (like the Bondi mass or angular momentum aspect) is provided, the solution will depend solely on $C_N$ and $C_S$.

\section{Taub-NUT without Closed Timelike Curves}

We now turn to study a full non-linear solution of the Einstein equations with a non-vanishing dual supertranslation charge - the Taub-NUT metric. The solution contains a string-like singularity. We start by reviewing Misner construction \cite{Misner:1963fr} of a singularity free solution which results in the appearance of \emph{Closed Timelike Curves} (CTC) and then turn to construct a singularity free solution without CTC along the lines described in the previous section.

The Taub-NUT metric is given by the following expression
\begin{equation}\label{TaubNUT}
ds^2=-f(r) \left(dt + 2 \ell \cos \theta d \varphi\right)^2 +\frac{dr^2}{f(r)} + \left(r^2 +\ell^2 \right) \left(d\theta ^2 +\sin^2 \theta d\varphi^2\right) ,
\end{equation}
with
\begin{equation}
f(r) = \frac{r^2 -2mr - \ell^2}{r^2 +\ell ^2},
\end{equation}
where $m$ is the mass aspect and $\ell$ is called the NUT parameter. The Taub-NUT metric has two horizons located at $r_{\pm}=m \pm \sqrt{m^2 +\ell ^2}$. The Taub region is defined by the domain $r_-<r<r_+$ while the NUT region is defined by the domain $r>r_+$ and $r<r_-$.

The Taub-NUT metric contains a string-like singularity along the axes $\theta=0$ and $\theta=\pi$. It is possible to partly remove the singularity and shift it around using diffeomorphisms. For example, using the change of coordinates $t \rightarrow t - 2\ell \varphi$ we can remove the singularity at $\theta=0$ while the metric transforms into
\begin{equation}\label{North}
ds_N^2=-f(r) \left(dt -4 \ell \sin^2 \frac{\theta}{2} d \varphi\right)^2 +\frac{dr^2}{f(r)} + \left(r^2 +\ell^2 \right) \left(d\theta ^2 +\sin^2 \theta d\varphi^2\right).
\end{equation}
The resulting metric contains a semi-infinite string singularity along the axis $\theta=\pi$. Similarly we can remove the singularity at $\theta=\pi$ by the change of coordinates $t \rightarrow t +2\ell \varphi$ applied to \eqref{TaubNUT}. The resulting metric is
\begin{equation}\label{South}
ds_S^2=-f(r) \left(dt +4 \ell \cos^2 \frac{\theta}{2} d \varphi\right)^2 +\frac{dr^2}{f(r)} + \left(r^2 +\ell^2 \right) \left(d\theta ^2 +\sin^2 \theta d\varphi^2\right),
\end{equation}
which contains a semi-infinite string singularity along the $\theta=0$ axis. It is not possible to remove completely the string singularity using diffeomorphisms, but only to change the direction of the semi-infinite string.

Misner constructed a solution without string singularities by dividing the spacetime into two patches, as described in the previous section. In the northern hemisphere he considered the metric \eqref{North} and in the southern hemisphere the metric \eqref{South}. As in the previous section, the metric is perfectly regular in each one of the patches separately. To prove that the solution is also differentiable in the overlap region, Misner argued that in this region the two metrics are equivalent up to a diffeomorphism given by
\begin{equation}\label{diff}
t_N = t_S +4 \ell \varphi.
\end{equation}
In other words, both the metric \emph{and} the spacetime manifold on which it is defined are divided into two patches in Misner's construction.
The consequence of \eqref{diff} is that since $\varphi$ is compact with a period of $2\pi$ then both $t_N$ and $t_S$ have to be compact with a period $8\pi \ell$ and the solution contains CTC.

What we would like to do now is to consider the same metrics on the two patches as Misner did but on the same, globally defined, spacetime manifold. Asymptotically the two metrics \eqref{North} and \eqref{South} are given by \eqref{bondiMetric} with
\begin{equation}
m_B=m
\end{equation}
and $C(z,\zb)$ given by $C_N$ and $C_S$ of \eqref{Cs}, respectively. As it was shown in the previous section, to leading order in the asymptotic expansion the two metrics are equivalent in the overlap region up to a regular 
dual supertranslation transformation. 
It is conceivable that the equivalence of the two metrics in the overlap region continue to hold to subleading orders, by virtue of subleading dual supertranslation gauge transformations \cite{Godazgar:2018dvh}. 

Our construction is analogous to the Wu-Yang monopole in QED, where a gauge transformation (and not a spacetime transformation, like in the Misner construction) is used to show the equivalence of the two asymptotic metrics in the overlap region. If the identification of metrics in the two hemispheres can be extended to the bulk, our construction gives a
 regular NUT solution with a metric that is defined using two patches, but on a globally defined spacetime manifold in which $t$ is a non-compact coordinate. However, we are still not guaranteed that there are no CTC, because the coordinate $\varphi$ is compact, so it could potentially become timelike. To examine this possibility that $\varphi$ becomes timelike we look at curves on which $t$, $r$ and $\theta$ are constant
\begin{equation}
\begin{aligned}
ds^2_N &= -\left[ 4\ell^2 f(r) (1-\cos\theta)^2 -(r^2 +\ell^2) \sin^2 \theta \right] d \varphi ^2,\\
ds^2_S &= -\left[ 4\ell^2 f(r) (1+\cos\theta)^2 -(r^2 +\ell^2) \sin^2 \theta \right] d \varphi ^2.
\end{aligned}
\end{equation}
These intervals are always spacelike when $f(r) \leq 0$, namely in the Taub part of the Taub-NUT metric. However, in the NUT part, where $f(r) \geq 0 $, they could possibly become timelike. In such a case, since $\varphi$ is a periodic coordinate, these intervals will form closed timelike curves.
In the following we will look for a region in the parameters space where CTC never form in the NUT region.

The coordinate $\varphi$ will become timelike on the northern and southern hemispheres when either of the following
inequalities hold
\begin{equation}\label{inequalities}
\begin{cases}
\begin{aligned}
	&\cos ( \pi- \theta) > \Delta(r) ,  							 \qquad  		    &     						0 \leq \theta  \leq \frac{\pi}{2} + \epsilon ,  \\
 &  \cos (  \theta)  > \Delta(r) ,   					 \qquad  			 & 						\frac{\pi}{2} - \ep  \leq \theta \leq \pi .
\end{aligned}
\end{cases}
\end{equation}
The function $\Delta(r) $ is given by the following ratio of polynomials
\begin{equation}
\Delta(r)  =  \frac{\Delta_-(r)}{\Delta_+(r)}, 
\qquad
\text{where}
\qquad
\Delta_{\pm} (r) = r^2 + \ell^2 \pm 4\ell^2 f(r) .
\end{equation}
We now take $\ep \rightarrow 0 $ such that the two patches are minimally overlapping.
In both patches, the left hand side of the inequalities \eqref{inequalities} is non-positive (in the corresponding domain).
The inequalities above therefore cannot be met if the right hand side --given in both cases by the function $\Delta(r)$-- is non-negative.
Since the function $\Delta_+ (r)$ is always positive in the NUT region, $\Delta(r)$ is non-negative if $\Delta_- (r)$ is non-negative.
The function $\Delta_-(r)$ is non-negative, in turn, when the bound \eqref{bound} is obeyed.
We therefore conclude that the inequalities above cannot be satisfied when the mass to NUT charge ratio obeys the bound \eqref{bound}, in which case no CTC will be formed.
The zero mass case is analogous to the magnetic monopole in QED.
The case when both $m$ and $\ell$ are non-zero is analogous to a dyon in QED.

 \section{Discussion}\label{ConclusionSection}

In this paper we have argued that dual supertranslation may be an Abelian gauge symmetry of asymptotically flat spacetimes by showing that all the S-Matrix elements are invariant under its action. Furthermore, we have shown that the Hilbert space of asymptotic states factorizes into distinct superselections sectors labeled by their supertranslation charges. The superselection rule then implies that states from different sector cannot transform into each other.
These results imply that general relativity may also possess an asymptotic gauge symmetry besides diffeomorphisms. It would be interesting to see if this gauge symmetry could be extended into the bulk of spacetime.

Next we explored solutions that belong to non-trivial superselection sectors. The Taub-NUT metric, which admits a  singularity analogous to the Dirac string in electrodynamics, provides such an example and was studied in~\cite{Kol:2019nkc}. In his seminal work \cite{Misner:1963fr}, Misner removes the string singularities from the Taub-NUT metric by constructing a monopole-like solution using two different patches that are identified on the equator by virtue of diffeomorphisms\footnote{It is interesting to note that Misner's paper \cite{Misner:1963fr} was published 13 years before the work of Wu and Yang \cite{Wu:1976ge}.}. However, Misner's construction results in the appearance of CTC, which prevent a physical interpretation of the solution. For more details on the vast literature related to Misner's work we refer the reader to the book \cite{Griffiths:2009dfa} and references therein. 
More recently, certain relations between magnetic solutions of general relativity and those of Yang-Mills theory were studied in \cite{Huang:2019cja,Luna:2015paa,Godazgar:2019ikr,Bahjat-Abbas:2020cyb}, but they did not shed new light on the problem of the CTC.
The main purpose of the work presented in this paper is to use dual supertranslation symmetry to construct a gravitational-magnetic monopole solution analogous to the Wu-Yang monopole solution in QED, which is regular everywhere on the celestial sphere and free of any string singularities. We further used the dual supertranslation symmetry to construct a non-linear Taub-NUT solution without strings and showed that when the bound \eqref{bound} is satisfied the solution is free of CTC that extend to null infinity.

The bound \eqref{bound} may remind the reader of the bound on angular momentum of the Kerr metric
\begin{equation}
\begin{aligned}
ds_{\text{Kerr}} ^2 &= - f  \Big(  dt - (1-f^{-1}) a \sin ^2 \theta d \phi \Big) ^2 
+\frac{\rho ^2}{\Omega} dr^2 +\rho ^ 2 \left(d\theta ^2 + \frac{\Omega}{f \rho^2} \sin ^2 \theta d\phi ^2\right) , \\
\rho^2 &= r^2 +a^2 \cos ^2 \theta ,\qquad
f = 1- \frac{2mr}{\rho^2}, \qquad
\Omega = r^2 -2mr +a^2.
\end{aligned}
\end{equation}
When $a \leq m$ the Kerr metric possesses two event horizons located at the zeros of the function $\Omega$
\begin{equation}
r_{\pm} = m \pm \sqrt{m^2 -a^2},
\end{equation}
but when $a>m$ there is a naked singularity. To avoid the appearance of a naked singularity the condition
\begin{equation}\label{boundKerr}
a\leq m
\end{equation}
must be imposed.
The condition \eqref{boundKerr} is an upper bound on the angular momentum while the new condition we found \eqref{bound} is a lower bound on the NUT charge, but otherwise they both seem to follow from the same principle,
which is in essence some form of cosmic censorship.

 Finally, let us comment on an alternative magnetic string solution that we can construct from the two patches \eqref{North} and \eqref{South}, by identifying antipodal points on the sphere as follows
 \begin{equation}
 \begin{aligned}
 \theta & \longleftrightarrow \pi - \theta, \\
 \varphi & \longleftrightarrow -\varphi .
 \end{aligned}
 \end{equation}
 This identification eliminates the discontinuity of the metric on the equator. The two fixed points of the antipodal map
 \begin{equation}
 \theta = \frac{\pi}{2}, \qquad \varphi = 0, \pi,
 \end{equation}
 remain singular, but the branch cut of the Taub-NUT string along the axis $\theta = 0,\pi$ is eliminated. The dual supertranslation charge of this solution is half the charge of the Taub-NUT string before the identification of anti-podal points. We present this construction as an example of a solution that carries a non-zero dual supertranslation charge,
 possesses string-like singularities on the celestial sphere, but does not have a branch cut along the whole string axis.

 \section*{Acknowledgements}
The research of UK and MP is supported in part by NSF through grant PHY-1915219.

 \appendix
 
 \section{The Magnetic Mode}\label{MagAppendix}
 
 The equations of motions describing the magnetic mode are given by eq.~\eqref{uzEquation}
 \begin{equation}\label{magneticEQS}
 	\begin{aligned}
 		\pa_u N_z &= +\frac{1}{4}\paz  \left(D_z^2 C^{zz} - D_{\zb}^2 C^{\zb\zb}\right)  -u \pa_u \paz m_B  - T_{uz} ,\\
 		\pa_u N_{\zb} &= - \frac{1}{4}\pazb  \left(D_z^2 C^{zz} - D_{\zb}^2 C^{\zb\zb}\right)  -u \pa_u \pazb m_B  - T_{u\zb} .
 	\end{aligned}
 \end{equation}
 We will use the Green's function to solve for the magnetic mode 
 \begin{equation}
 	\ImCzz \equiv \frac{D_z^2 C^{zz} - D_{\zb}^2 C^{\zb\zb}}{2i}.
 \end{equation}
 The Green's function obeys
 \begin{equation}
 	\paz \pazb G(z;w) = -  \paz \pa_{\wb} G(z;w) = -  \pazb \pa_{w} G(z;w) = \delta^2 (z-w).
 \end{equation}
 The solution for the Green's function is
 \begin{equation}
 	G(z;w) =  \frac{1}{2 \pi} \log |z-w|^2
 \end{equation}
 and it obeys
 \begin{equation}
 	\begin{aligned}
 		\paz G(z;w) &= -\pa_w G(z;w) =\frac{1}{2 \pi}  \frac{1}{z-w}, \\
 		\pazb G(z;w) &= -\pa_{\wb} G(z;w) =\frac{1}{2 \pi}  \frac{1}{\zb-\wb}.
 	\end{aligned}
 \end{equation}

 We can now solve both equations \eqref{magneticEQS}
 \begin{equation}\label{magSol1}
 	\begin{aligned}
 		\frac{i}{2}\ImCzz &= + \int d^2 w \, \pa_w G(z;w) \left( \pa_u N_{\wb} +T_{u\wb}\right) + \int d^2w \,   \pa_w G(z;w)  \, u \pa_u  \pa_{\wb} m_B \\
 		&= -  \int d^2 w \, \pa_{\wb} G(z;w)  \left( \pa_u N_{w} +T_{u w}\right) - \int d^2w \, \pa_{\wb} G(z;w) \,  u \pa_u \pa_w m_B.
 	\end{aligned}
 \end{equation}
 The two solutions have to be equal for consistency. Let's see how it works. Consider the integral
 \begin{equation}
 	\int d^2w \,   \pa_w G(z;w)  \, u \pa_u  \pa_{\wb} m_B =
 	-\int d^2w \,   G(z;w)  \, u \pa_u \pa_w  \pa_{\wb} m_B +\int d^2w \,   \pa_w \Big ( G(z;w)  \, u \pa_u  \pa_{\wb} m_B \Big).
 \end{equation}
 We now assume that the Bondi mass is regular on the sphere, so the second term in the equation above vanishes, since it is  an integral over a total derivative. We can now rewrite \eqref{magSol1}
 \begin{equation}\label{magSol2}
 	\begin{aligned}
 		i \pi \ImCzz &= + \int d^2 w \, \pa_w G(z;w) \left( \pa_u N_{\wb} +T_{u\wb}\right) - \int d^2w \,  G(z;w)  \, u \pa_u  \pa_w  \pa_{\wb} m_B \\
 		&= -  \int d^2 w \, \pa_{\wb} G(z;w)  \left( \pa_u N_{w} +T_{u w}\right) + \int d^2w \,  G(z;w) \,  u \pa_u \pa_w \pa_{\wb} m_B.
 	\end{aligned}
 \end{equation}
 From the second equality in \eqref{magSol2} we can deduce that
 \begin{equation}
 	\int d^2w \, G(z;w) \, u \pa_u \pa_w \pa_{\wb} m_B  = \frac{1}{2}
 	\int d^2 w \, 
 	\Big(
 	\pa_w G(z;w) \left( \pa_u N_{\wb} +T_{u\wb}\right)
 	+
 	\pa_{\wb} G(z;w)  \left( \pa_u N_{w} +T_{u w}\right)
 	\Big).
 \end{equation}
 Plugging back into the solution for $\ImCzz$ we now have
 \begin{equation}\label{solMag}
 	\ImCzz 
 	= i  \int d^2 w \, 
 	\Big(
 	\pa_{\wb}G(z;w) \left( \pa_u N_{w} +T_{u w}\right)
 	-
 	\pa_{w}G(z;w) \left(  \pa_u N_{\wb} +T_{u \wb}\right)
 	\Big).
 \end{equation}

 Let us now comment on a particular subtlety. The equations of motion \eqref{magneticEQS} can be manipulated by taking derivatives in order to write an explicit equation for the magnetic mode
 \begin{equation}\label{simEq}
 	\pa_u \pa_{[\zb} N_{z]} = i \paz \pazb \Im \left[  D_z^2 C^{zz}(u,z,\zb) \right]  - \pa_{[\zb}T_{uz]}.
 \end{equation}
 The last equation can be solved by
 \begin{equation}\label{notSol}
 	\ImCzz=-
 i\int d^2 w \, G(z;w) \, 
 	\Big(
 	\pa_u \pa_{[\wb}N_{w]} + \pa_{[\wb}T_{w]u}
 	\Big).
 \end{equation}
 The expression in \eqref{notSol} differs from the solution \eqref{solMag} by a boundary term
 \begin{equation}
 	i \int d^2 w \, 
 	\Big[
 	\pa_w
 	\Big(
 	G(z;w) \, 
 	\left(
 	\pa_u  N_{\wb} + T_{u \wb }
 	\right)
 	\Big)
 	-
 	\pa_{\wb}
 	\Big(
 	G(z;w) \, 
 	\left(
 	\pa_u  N_{w} + T_{u w }
 	\right)
 	\Big)
 	\Big].
 \end{equation}
 This boundary term vanishes if $\pa_u  N_{w} + T_{u w }$ is regular on the sphere; however, it does not vanish 
 when the stress tensor is not regular on the sphere.  We would like to emphasize that this boundary term is not part of the solution. It arises only because we have taken extra derivatives in order to write the equation \eqref{simEq}. These extra derivative were inverted, in turn, and produced these boundary terms. However, this is an artifact that is not part of the solution of the original equations of motion.


\end{document}